\documentclass[12pt]{article}
\usepackage[mathscr]{eucal}
\usepackage{amsfonts}
\usepackage{amssymb,latexsym}
\usepackage{graphics}
\usepackage{theorem}

\makeatletter
\renewcommand{\section}{\@startsection{section}{1}{0mm}%
{-0.5\baselineskip}{1 pt}{\normalfont\large\bfseries}}
\renewcommand{\subsection}{\@startsection{subsection}{2}{0mm}%
{-0.1\baselineskip}{1 pt}{\normalfont\bfseries}}
\renewcommand{\subsubsection}{\@startsection{subsubsection}{3}{0mm}%
{-0.1\baselineskip}{1 pt}{\normalfont\bfseries}}
\renewcommand{\paragraph}{\@startsection{paragraph}{4}{0mm}%
{-0.1\baselineskip}{1 pt}{\normalfont\itshape}}
\renewcommand{\subparagraph}{\@startsection{subparagraph}{5}{0mm}%
{-0.1\baselineskip}{1 pt}{\normalfont}}
\makeatother

\newcommand{\RR}{\mathbb{R}}
\newcommand{\ZZ}{\mathbb{Z}}
\newcommand{\TT}{\mathbb{T}}
\newcommand{\SSS}{\mathbb{S}}
\newcommand{\HH}{\mathbb{H}}

\theoremstyle{plain} 
\theorembodyfont{\normalfont\slshape}

{\theorembodyfont{\normalfont\slshape}}

\newcommand{\SLR}{\widetilde{\textup{SL}}(2,\mathbb{R})}

\def\endproof{\nobreak\kern5pt\nobreak\vrule height4pt width4pt depth0pt}

\begin{document}
\begin{titlepage}
\vspace*{1cm}
\begin{center}
{\Large \bf New Five Dimensional Black Holes Classified by Horizon Geometry, 
and a Bianchi VI Braneworld}\\ 
\vspace*{2cm}
C.\ Cadeau\footnote
{e-mail: ccadeau@phys.ualberta.ca. Address after 1 Sept., 2000: 
Dept.\ of Physics, 
University of Alberta, Edmonton, AB, Canada T6G~2J1.}
and E.\ Woolgar\footnote{e-mail: ewoolgar@math.ualberta.ca}\\
\vspace*{0.5cm}
Dept. of Mathematical Sciences, CAB 632\\
and\\
Theoretical Physics Institute\\
University of Alberta\\
Edmonton, AB\\
Canada T6G\ 2G1\\
\vspace*{2cm}

\begin{abstract}
\noindent 
We introduce two new families of solutions to the vacuum Einstein
equations with negative cosmological constant in $5$ dimensions. 
These solutions are static black holes whose horizons are modelled 
on the $3$-geometries nilgeometry and solvegeometry. Thus the horizons 
(and the exterior spacetimes) can be foliated by compact $3$-manifolds 
that are neither spherical, toroidal, hyperbolic, nor product manifolds, 
and therefore are of a topological type not previously encountered in 
black hole solutions. As an application, we use the solvegeometry solutions 
to construct Bianchi VI$_{-1}$ braneworld cosmologies.
\end{abstract}
\end{center}
\end{titlepage}

\setcounter{equation}{0}
\section{Introduction}
In 1972, Hawking \cite{Ha1} proved the first black hole topology theorem, 
which stated that the smooth event horizon of a stationary black hole 
obeying the dominant energy condition was spherical. Hawking followed 
this with a topology theorem for apparent horizons \cite{Ha2}, while 
Gannon \cite{Ga1} was able to remove the assumption of stasis in favour 
of a certain regularity condition. Much later, Chru\'sciel and Wald 
\cite{ChWa1} were able to apply the active topological censorship 
theorem to obtain a new topology theorem for black holes in 
asymptotically flat spacetimes that did not rely on smoothness 
of the event horizon.

Nonetheless, within the last decade, it has become clear that under 
appropriate circumstances the event horizon of a 4-dimensional black 
hole can be a compact Riemann surface of arbitrary genus. Examples 
are provided by the various black holes discovered in locally anti-de 
Sitter backgrounds \cite{AmBeHoPe1,Ma1,Le1}. These black holes avoid 
Hawking's theorem because they do not obey the dominant energy condition. 
Hawking's basic {\it argument} still applies in the absence of this 
condition, but leads only to a lower bound on the area of the horizon, 
not to a restriction on the topology \cite{Wo1}. Topological censorship 
arguments also apply, and they lead to a {\it genus inequality} stating 
that the genus of the horizon or, more generally, the sum of the genera 
of all components of the horizon, cannot exceed the genus of scri 
\cite{GaScWiWo1}.\footnote
{From this, one recovers the result of Chru\'sciel and Wald as the 
special case in which the genus of scri is zero. Then the
genus inequality asserts that the components of the horizon must 
be spherical. 

There are also so-called temporarily toroidal black holes that occur 
in numerical simulations of gravitational collapse of asymptotically 
flat initial data. These horizons cut certain spacelike hypersurfaces 
in a torus, but cut later hypersurfaces in a sphere. The phenomenon is 
a consequence both of the nature of the collapse and of the choice of 
spacelike hypersurfaces. For the explanation of these horizons in 
the context of topological censorship, see \cite{GaScWiWo1}. Moreover, 
the discussion in \cite{GaScWiWo1} leads one to conclude that toroidal 
horizons of this nature will always contain ``crossover points'' where 
generators of the horizon begin, and so will not be smooth, thus 
circumventing Hawking's theorem.}
Thus, black holes with non-zero genus horizons have boundaries-at-infinity 
(scri) that do not have the usual topology $\SSS^2\times \RR$ for anti-de 
Sitter space. 

Topological censorship constraints on spacetime topology have 
a surprising relevance to the case of $5$-dimensional spacetime, 
where they resolve a puzzle that would otherwise plague the AdS/CFT 
correspondence \cite{GaScWiWo2,WiYa1}. Now in $5$ dimensions, the 
event horizon of a stationary black hole is a ruled hypersurface 
foliated by a compact orientable $3$-dimensional Riemannian manifold, 
dragged along the null generators. Thurston's famous geometrization 
conjecture, if correct, would provide a classification of such 3-manifolds, 
and hence of horizon topologies in $5$ dimensions. The conjecture 
asserts that $3$-dimensional compact manifolds can be decomposed 
by cutting along certain embedded $2$-spheres and incompressible tori 
in a unique way such that the resulting pieces are each covered by 
exactly one ``model geometry.'' We define this notion more precisely 
in Section 2.1. Thurston has shown that there are eight model geometries. 
The model geometries admit homogeneous metrics and are very nearly in 
correspondence with the nine homogeneous geometries of Bianchi. 
Descriptions of the model geometries and Thurston's work can be found 
in Thurston's book \cite{Th1} and the article by Scott \cite{Sc1}. The 
correspondence with Bianchi models is detailed in \cite{Fa1}.

Then a natural question is whether static black hole horizons in 
$5$ dimensions can be built from arbitrary compact 3-manifolds.
A less ambitious problem is to begin with the Thurston model
geometries themselves, and ask, ``For each Thurston model geometry, 
can one find a static $5$-dimensional Einstein manifold with a black 
hole whose event horizon is foliated by a compact $3$-manifold modelled 
on the given model geometry?'' In the present work we explore this issue.

The question is of interest from several points of view. First, 
if it proves difficult to construct solutions for various horizon
topologies, then this may indicate that there are new constraints on
horizon topology. Such constraints may even descend to $4$ dimensions.
Second, in $4$ and $5$ dimensions, the search for $g>0$ black holes 
has led to spacetimes that exhibit unexpected and remarkable properties,
including negative mass solutions that are not nakedly singular, 
being either black holes or geodesically complete. This has led Myers 
and Horowitz to conjecture a new positive energy theorem for these 
solutions in which negative but bounded mass is permitted \cite{Gi1,MyHo1}. 
It would be interesting to determine whether similar behaviour occurs 
when more general topologies are permitted. Third, present ideas in 
dimensional reduction suggest that our cosmos may be a $3$-brane evolving
in a $5$-dimensional spacetime. As we will see below, in pursuing the 
above question, we will be led to homogeneous but non-FRW braneworld 
cosmologies.

Before one can address such issues, it is important to have available 
exact solutions from which to develop intuition and against which to 
test hypotheses. Therefore, in the present article, we focus primarily 
on the question of whether we can obtain solutions with horizons modelled 
on the Thurston geometries. We outline our approach to this in Section 2.1. 
For the $5$ ``untwisted'' model geometries, such black holes are easy to 
obtain and have already appeared in the literature; we briefly discuss 
these cases in Section 2.2. In Sections 2.3 and 2.4, we give two new 
families of black hole solutions whose horizons are modelled on the Sol 
and Nil $3$-geometries, respectively. These solutions admit topologies 
not previously encountered in black holes. We give explicit constructions 
of some of these topologies. The case of a horizon modelled on the 
$\SLR$ Thurston geometry remains open. We illustrate one 
application of our new solutions: from the solvegeometry black hole 
we find Bianchi VI$_{-1}$ braneworld cosmologies in Section 3.

Throughout this article, early roman indices are abstract indices. Middle
roman indices label elements of a triad of basis vectors and indicate
components of tensors with respect to that basis. We assume all manifolds
are Hausdorff.

\setcounter{equation}{0}
\section{Model $3$-Geometries and Black Hole Solutions}
\subsection{Preliminaries}
A \emph{model geometry} is defined as a pair $(X,G)$, where $X$ is 
a connected and simply connected $n$-manifold and
\begin{itemize}
\item[({\it i})]  $G$ is a Lie group of
diffeomorphisms acting transitively on $X$ with compact point 
stabilisers,\footnote
{The {\it stabiliser} of $p\in X$ is the isotropy group at $p$.}
\item[({\it ii})] $G$ is maximal; {\it i.e.}, $G$ is not a proper 
subgroup of a larger group $H$ acting in the required way on $X$, and
\item[({\it iii})] there is a subgroup $\Gamma \leq G$ acting on $X$ as a 
covering group, such that the quotient $M$ is a compact $n$-manifold.
%
%
\end{itemize}
Any $M$ meeting the requirements of point ({\it iii}) is said to be 
{\it modelled on} $(X,G)$, and is called an $(X,G)$ manifold. Note 
that point ({\it i}) implies that $X$ admits a complete, homogeneous, 
Riemannian metric invariant by $G$ ({\it cf.}\ propositions 3.4.10 and 
3.4.11 of \cite{Th1}), a fact that we will exploit. 

There are $8$ model $3$-geometries (\cite{Th1}, Theorem 3.8.4).
They are denoted as spherical (or elliptical), 
hyperbolic, Euclidean (or flat), $\SSS^2\times \SSS^1$, $\HH^2\times \SSS^1$, 
nilgeometry, solvegeometry, and $\SLR$. 
While manifolds modelled on the first five geometries are familiar, 
manifolds modelled on last $3$, the so-called ``twisted product'' 
cases, may not be to many readers. Sections 2.3 and 2.4 therefore 
contain explicit constructions of compact Sol- and Nil-manifolds, 
respectively. Further detail can be found in \cite{Th1,Sc1}. We 
will not deal with $\SLR$ manifolds 
in what follows, except for brief remarks in the Discussion section.

Our strategy is first to find 5-dimensional spacetimes $(M,g_{ab})$
foliated by spacelike 3-surfaces $X_{t,r}$ that carry one of the 
3-geometries $(X,G)$. We assume a product topology, for several reasons.
This assumption will not only lead to relatively simple Einstein 
equations, but will also ensure we do not come up against the known
topological censorhip constraints, which can forbid some non-product 
topologies. Moreover, we seek spacetimes with a null hypersurface as 
inner boundary, representing an event horizon, and by the product 
topology assumption this boundary will also be foliated by the 
relevant $3$-geometry. We further assume that
\begin{itemize}
\item[({\it i})] the spacetime is a static Einstein manifold whose
\item[({\it ii})] time-symmetric hypersurfaces are foliated by 
homogeneous $3$-surfaces generated by isometries of the spacetime.
\end{itemize}
We have chosen to impose assumption ({\it ii}) here because it 
considerably simplifies the second part of our task, which is 
to determine compact topologies for the surfaces $X_{t,r}$. Each 
model geometry $(X,G)$ admits a homogeneous metric invariant by $G$, 
so this assumption implies that $G$ is a subgroup of the spacetime 
isometry group, with orbits that are $3$-surfaces foliating spacetime, 
such that the induced metric on the orbits is $G$-invariant. We then 
seek to identify spacetime points that are related by isometries in 
$G$ acting freely and properly discontinuously. Since the identifications 
are by isometries, the quotient metric will be smooth. Our task will be to 
choose the isometries to have cocompact action on each leaf.\footnote
{An action of $\Gamma$ on a locally compact space $X$ is {\it properly
discontinuous} iff for every compact subset $C\subseteq X$ the set 
$\{\gamma\in\Gamma:\ \gamma C\cap C\neq\emptyset\}$ is finite. It is
{\it free} if $\gamma x \neq x$ whenever $\gamma\in\Gamma$ is not the 
identity. If $\Gamma$ acts properly discontinuously on $X$, it is called 
a {\it discrete group of transformations of $X$}, or more simply a 
{\it discrete group}. If a discrete group $\Gamma$ acts freely on $X$, 
then $X/\Gamma$ will be a (Hausdorff) manifold. If $X/\Gamma$ is compact, 
the action of $\Gamma$ (or $\Gamma$ itself) is called {\it cocompact}.} 

Under the above conditions, one can write the metric in the form 
\begin{equation}
ds^2=-V(r)dt^2+\frac{dr^2}{V(r)}+h_{ij}(r)\omega^i\omega^j\quad ,
\label{eqII1}
\end{equation}
where the $\omega^i$ are invariant 1-forms and $h_{ab}(r)=h_{ij}(r)
\omega^i_a\omega^j_b$ is the metric induced on the $t=const.$, $r=const.$
surfaces. Our invariant 1-forms are those found in Ryan and Shepley 
\cite{RySh1}. The horizon is the zero set of $V(r)$. The requirement that 
(\ref{eqII1}) be an Einstein metric can be written in dimension $n$ as
\begin{equation}
R_{ab}=\frac{2\Lambda}{n-2}g_{ab}\quad ,\label{eqII2}
\end{equation}
where $R_{ab}$ is the Ricci tensor and $\Lambda$ is the cosmological 
constant. 

\subsection{Constant Curvature Horizons and Product Horizons}
Here we catalogue solutions whose horizon topologies are modelled 
on $3$-geometries of type spherical, flat, hyperbolic, $\HH^2\times 
\SSS^1$ (where $\HH^2$ is $2$-dimensional hyperbolic space), and 
$\SSS^2\times \SSS^1$.

Consider an $n$-dimensional warped product spacetime
\begin{equation}
(M,g_{ab})=({\overline M},{\overline g}_{ab})\times_{f^2}
({\tilde M},{\tilde g}_{ab})\label{eqII3}\quad ,
\end{equation}
with metric ${\overline g}_{ab}\oplus f^2{\tilde g}_{ab}$, 
$f:{\overline M}\to \RR$, where $({\overline M},{\overline g}_{ab})$ 
is a spacetime of dimension ${\overline n}$ and $({\tilde M},
{\tilde g}_{ab})$ is a Riemannian manifold of dimension ${\tilde n}$, 
so ${\rm dim}(M)=:n={\overline n}+{\tilde n}$. We use bars to denote 
$({\overline M},{\overline g}_{ab})$ quantities, tildes to denote 
$({\tilde M},{\tilde g}_{ab})$ quantities, and no adornment to denote 
$(M,g_{ab})$ quantities in what follows. The Ricci curvature of such 
a product is then the direct sum of two terms:
\begin{eqnarray}
R_{ab}&=& \left [ {\overline R}_{ab}-\frac{{\tilde n}}{f(r)}
{\overline \nabla}_a {\overline \nabla}_b f(r) \right ] 
\nonumber\\
&&\oplus \left [ {\tilde R}_{ab}-{\tilde g}_{ab}\left ( f(r) 
{\overline \Delta}f(r) +({\tilde n}-1){\overline g}^{ab}
{\overline \nabla}_a f(r) {\overline \nabla}_b f(r)\right )
\right ]\quad ,\label{eqII4}
\end{eqnarray}
where ${\overline \nabla}_a$ and ${\overline \Delta}$ are 
respectively the covariant derivative and d'Alembertian on 
$({\overline M},{\overline g}_{ab})$. When $g_{ab}$
is an Einstein metric, ${\tilde R}_{ab}$ will satisfy an 
equation of the form ${\tilde R}_{ab}+{\tilde g}_{ab}(\dots)=0$. 
When ${\tilde n}>2$, the contracted Bianchi identity on $({\tilde M},
{\tilde g}_{ab})$ implies that the coefficient of ${\tilde g}_{ab}$ 
must be constant and therefore ${\tilde g}_{ab}$ will necessarily 
be an Einstein metric on ${\tilde M}$.

Taking $i,j\in{{\overline n},\dots,n-1}$ and specializing to 
${\overline n}=2$, we take the line elements of these metrics to be
\begin{eqnarray}
ds^2&=&d{\overline s}^2 + f^2(r)d{\tilde s}^2\label{eqII5}\quad ,\\
d{\overline s}^2&=&-V(r)dt^2 + \frac{dr^2}{V(r)}\label{eqII6}\quad ,\\
d{\tilde s}^2&=&{\tilde g}_{ij}dx^idx^j\label{eqII7}\quad .
\end{eqnarray}
By using (\ref{eqII5}--\ref{eqII7}) to expand (\ref{eqII4}) and substituting 
the results into the Einstein equations (\ref{eqII2}), we obtain
\begin{eqnarray}
&&\frac{1}{2}V''(r)+\frac{(n-2)}{2}V'(r)\frac{f'(r)}{f(r)}
=-\frac{2}{n-2}\Lambda\label{eqII8}\quad ,\\
&&f''(r)=0\label{eqII9}\quad ,\\
&&\frac{1}{f^2(r)}{\tilde R}_{ij}=\left [ (n-3)V(r)\left (
\frac{f'(r)}{f(r)}\right )^2+V'(r)\frac{f'(r)}{f(r)}
+\frac{2}{n-2}\Lambda\right ] {\tilde g}_{ij}\label{eqII10}\quad .
\end{eqnarray}

Solving these equations, we find (up to constant rescalings and 
translations of the $r$ coordinate) two classes of solutions, 
depending on the integration of $f''(r)$. The first is
\begin{eqnarray}
&&f(r)=r\label{eqII11}\quad ,\\
&&V(r)=-\frac{2\Lambda}{(n-1)(n-2)}r^2 + k -\frac{2M}{r^{n-3}}
\label{eqII12}\quad ,\\
&&{\tilde R}_{ij}=(n-3)k\delta_{ij}\label{eqII13}\quad ,
\end{eqnarray}
where $k$ and $M$ are constants, and $({\tilde M},{\tilde g}_{ab})$
is an Einstein $(n-2)$-manifold of scalar curvature $(n-2)(n-3)k$. 
In the $n=5$ case, the solutions of (\ref{eqII13}) are the constant 
curvature metrics in dimension $3$:
\begin{equation}
d{\tilde s}^2=
\left \{ 
\begin{array}{ll}
d\theta^2+\sin^2\theta\left ( d\phi^2 +\sin^2\phi\ d\xi^2\right )\ ,
&k=1\ ,\\
d\theta^2+d\phi^2+d\xi^2\ ,&k=0\ ,\\
d\theta^2+\sinh^2\theta\left ( d\phi^2 +\sin^2\phi\ d\xi^2\right )\ ,
&k=-1\ .\label{eqII14}
\end{array}
\right .
\end{equation}
These solutions appear in Birmingham \cite{Bi1}, and include 
higher-dimensional versions of the Schwarzschild solution 
($\Lambda=0$, $k=1$), Kottler (also called ``AdS-Schwarzschild,'' 
$\Lambda<0$, $k=1$), ``hyperbolic Kottler'' ($\Lambda<0$, $k=-1$) 
\cite{AmBeHoPe1,Ma1}, and Lemos ($\Lambda<0$, $k=0$) \cite{Le1} 
solutions. (When $n>5$, there are solutions of (\ref{eqII13}) that 
are not constant curvature metrics; {\it cf.}\ \cite{Go1} for an 
$n=6$ metric foliated by $\SSS^2\times \SSS^2$.)

For Kottler and Schwarzschild solutions, the $t=const.$, $r=const.$
surfaces will carry spherical geometry. They can have the topology 
of $\SSS^{n-2}=\SSS^{{\tilde n}}$ or its quotient by a discrete group of 
isometries, such as the lens spaces or, for ${\tilde n}=3$, Poincar\'e 
dodecahedral space (see Theorem 4.4.14 of \cite{Th1} for a classification 
of all possibilities when ${\tilde n}=3$). For the hyperbolic Kottler 
solutions, the surfaces $t=const.$, $r=const.$ can be compact hyperbolic 
${\tilde n}$-manifolds. In the Lemos case,\footnote
{Starting from the Kottler solution, one can obtain both the 
hyperbolic and the Lemos solutions without further reference 
to the field equations. For example, in the $n=5$ case, to 
obtain hyperbolic solutions, simply make the replacements $t\to it$, 
$r\to ir$, $\theta\to i\theta$, and $M\to iM$ in the Kottler solution.
To obtain the Lemos solutions, we follow Witten \cite{Wi1}, who 
worked with a Euclidean version of Lemos's metric. In the Kottler 
solution, in arbitrary dimension, let $r=(M/\mu)^{1/(n-1)}\rho$, 
$t=(\mu/M)^{1/(n-1)}\tau$. Then the metric takes the form $-W(\rho) 
d\tau^2+ d\rho^2/W(\rho) + (M/\mu)^{2/(n-1)}\rho^2 d\Omega^2$ where 
$W(\rho)=(\rho/\ell)^2+k(\mu/M)^{2/(n-1)}-2\mu/\rho^{n-3}$, $\ell^2=
(n-1)(n-2)/(-2\Lambda)$, and $d\Omega^2$ is the round sphere metric. 
Now take $M\to\infty$. Then $W(\rho)\to (\rho/\ell)^2-2\mu/\rho^{n-3}$ 
and the radius of curvature of the induced metric $(M/\mu)^{2/(n-1)}\rho^2 
d\Omega^2$ on the $\rho=const$, $\tau=const$ submanifolds becomes infinite, 
so it goes over to the flat metric. Then the full metric takes the Lemos 
form with mass parameter $\mu$. One might therefore think of the Lemos 
metric as being separated from the Kottler ones by an infinite mass gap. 

In light of these simple tricks, it is surprising the Lemos and hyperbolic 
Kottler solutions were not discovered until over $75$ years after Kottler's 
work (a class of toroidal horizons was reported in 1979 \cite{Pe1}). Four 
years after Lemos's paper, the $5$-dimensional Lemos metric first appears 
in \cite{HoRo1}, as a dimensional reduction of a $10$-dimensional $3$-brane 
metric. The charged Lemos metric appeared in \cite{HuLi1}. Both these 
results are independent rediscoveries/generalizations; neither seem to 
have been aware of \cite{Le1}. Many subsequent string theory papers trace 
this metric back only as far as \cite{HoRo1}.}
these surfaces are closed flat manifolds. These are quotients of 
$\TT^2\times \RR$ by discrete groups; {\it e.g.}, $\TT^3$.

The second class of solutions is given by
\begin{eqnarray}
f(r)&=&1\label{eqII15}\quad ,\\
V(r)&=&-\frac{2}{3}\Lambda r^2 + C\label{eqII16}\quad ,\\
{\tilde R}_{ij}&=&\frac{2}{n-2}\Lambda\delta_{ij}\label{eqII17}\quad .
\end{eqnarray}
These solutions are products of a $2$-dimensional Einstein spacetime 
with an Einstein $(n-2)$-manifold, with no warping. They have recently
appeared in \cite{CaVaZe1} in connection with the near-horizon 
approximation. This class contains the Nariai solution as the $n=4$, 
$\Lambda>0$ case. These solutions may also be thought of as cosmological 
analogues of the vacuum case of the Bertotti-Robinson solutions. 
They admit the same topologies as the Birmingham solutions, but do
not have a scri (essentially because various metric coefficients 
are not $\sim r^2$). That is, they admit a notion of conformal infinity 
sufficient to define the event horizon as the boundary of the region 
in causal contact with it, but conformal infinity has codimension $>1$, 
and so is not a boundary-at-infinity. 

Other analogues of the Nariai and Bertotti-Robinson solutions can be 
obtained by again choosing the warping function $f=1$ in (\ref{eqII4}), 
but this time we take ${\overline n}={\rm dim}({\overline M})=3$. 
Then the Ricci curvature decomposes into a direct sum of the Ricci
curvature on a $3$-dimensional manifold and the Ricci curvature on an
$(n-3)$-dimensional one. Thus, to obtain a $5$-dimensional solution 
with negative cosmological constant and the horizon modelled on the 
$3$-geometry $\HH^2\times \SSS^1$, we merely take the BTZ black hole as 
our $3$-manifold and any constant (negative) curvature $g>1$ Riemann 
surface as the other factor. The metric is
\begin{equation}
ds^2 = -\left ( \frac{-\Lambda}{3}r^2 -M\right ) dt^2 + \frac{dr^2}
{\frac{-\Lambda}{3}r^2-M} + r^2 d\xi^2 + \frac{3}{(-2\Lambda)}\left (
d\theta^2 + \sinh^2 \theta d\phi \right )\label{eqII18}\quad ,
\end{equation}
where $\Lambda<0$ and so we have a horizon whenever the BTZ mass 
constant $M$ is positive. 

To obtain a black hole of horizon topology $\SSS^2\times \SSS^1$, one might
try replacing $\theta$ by $i\theta$, with the effect that $\sinh\theta$
is replaced by $\sin\theta$ and $\Lambda$ must now be $>0$ in (\ref{eqII18}). 
Letting $\mu=-M$, we obtain that the coefficient of $dt^2$ is $>0$, and 
that of $dr^2$ is $<0$, iff $r^2<3\mu/\Lambda$, so this is not a black 
hole exterior solution. However, an example of a black hole that admits 
this topology (with zero cosmological constant) is the black string obtained 
by appending a $d\xi^2$ term to the $4$-dimensional Schwarzschild solution.
\footnote
{One can understand the absence of ``unwarped'' product solutions 
${\overline M}^3\times \SSS^2$ generally. By (\ref{eqII17}), the presence 
of the $\SSS^2$ factor implies $\Lambda>0$. But recent work of Ida \cite{Id1} 
shows that ${\overline M}^3$ will be a black hole only if $\Lambda<0$, a 
contradiction.}

\subsection{Solvegeometry Black Holes}
We turn now to the new solutions with twisted product horizons. The first 
case is that of solvegeometry.

The $3$-manifold $X$ upon which solvegeometry manifolds are modelled is 
the Lie group Sol, described by the semidirect product $\RR^{2} \rtimes 
\RR$ with the multiplication given by
\begin{equation}
((x,y),z)\cdot((x',y'),z') = ((x+e^{-z}x',y+e^{z}y'),z+z')\quad .
\label{eqII19}
\end{equation}
The Sol-invariant $1$-form fields $\omega^1=e^zdx$, $\omega^2=e^{-z}dy$, and 
$\omega^3=dz$ can be used to construct a (left) invariant metric on Sol 
of Bianchi type VI$_{-1}$~\cite{Fa1}. The identity component of the 
isometry group for Sol is just Sol itself.

We have found the following family of solutions of the Einstein equation 
(\ref{eqII2}) with cosmological constant $\Lambda<0$:
\begin{eqnarray}
ds^2 &=& -\left ( \frac{-2\Lambda}{9}r^2 - \frac{2M}{r}\right ) dt^2 +
\frac{dr^2}{\frac{-2\Lambda}{9}r^2 - \frac{2M}{r}}
+\frac{3}{-\Lambda} d{\tilde s}^2\label{eqII20}\quad ,\\
d{\tilde s}^2 &=& r^2 \left (e^{2z} dx^2 + e^{-2z} 
dy^2 \right )+dz^2 \label{eqII21}\quad .
\end{eqnarray}
For any positive value of the integration constant $M$, there is 
one horizon, located at $r=(9M/(-\Lambda))^{1/3}$. We may regard 
$d{\tilde s}^2$ as giving the induced metric on $r=const.$, 
$t=const.$\ surfaces. This induced metric is of Bianchi type 
VI$_{-1}$ and so is Sol-invariant. We can extend the action of Sol 
on these surfaces to an action of Sol on the spacetime, such that
the orbits are these surfaces. 

Because we may take these orbits to be copies of $\RR^3$, these 
solutions can be regarded as ``black $3$-branes.'' To construct 
black holes, we now compactify the orbits, making them compact 
$3$-manifolds modelled on solvegeometry. We start with a $2\times 2$ 
symmetric matrix $M$ other than the identity, having unit determinant, 
positive trace, and integer entries. These conditions are not overly 
restrictive, and amount to finding the solutions of $ab=1+c^2$ in
positive integers; {\it e.g.}, $\left [\begin{array}{cc}2&1\\ 1&1 
\end{array}\right ]$ and $\left [ \begin{array}{cc}2&3\\ 3&5 \end{array}
\right ]$ are examples. Any such matrix preserves the lattice $\ZZ^2
\subseteq\RR^2$. 

Now $M$ is defined to have orthogonal eigenvectors and distinct, real,
reciprocal eigenvectors which we may write as $e^{\pm a}$, $a\neq 0$.
Align the $x$- and $y$-axes in $\RR^3$ to lie along the two eigenspaces. 
Then we may define
\begin{equation}
\psi(x,y,z)=(M\cdot(x,y),z+a)=(e^{-a}x, e^{a}y, z+a)\quad .\label{eqII22}
\end{equation}
From (\ref{eqII19}), $\psi$ is the action of the element $(0,0,a)$ of
Sol. Finally, define two independent translations $T_{1,2}:\RR^3\to
\RR^3:(x,y,z)\mapsto (x',y',z)$, both preserving $z$, and preserving
the same lattice as $M$ does (being careful to note that this lattice 
is the set of points with integer coordinates in the {\it original} 
basis, not the eigenvector basis, a source of some confusion in a similar 
discussion on p.389 of \cite{Th1} ). By (\ref{eqII19}), these are also 
Sol transformations. It is now an easy exercise to show that the Sol 
subgroup generated by $(0,0,a)$ and the two translations is discrete
(meaning that it acts properly discontinuously) and acts freely. 
Therefore, the quotient of $\RR^3$ by this subgroup is a Sol 
manifold.\footnote
{Another viewpoint is obtained by first identifying points in $\RR^3$
under the two translations, to obtain a ``toroidal cylinder.'' Then 
$\psi$ induces an automorphism $\phi$ of the torus, which can be used 
to glue the two toroidal ends of the cylinder together. This construction 
is called the {\it mapping torus} of $\phi$.}
With the natural projection $(x,y,z)\mapsto z\in S^1$, it has the 
structure of a torus bundle over the circle, and so is compact.
\begin{figure}
\begin{center}
\includegraphics{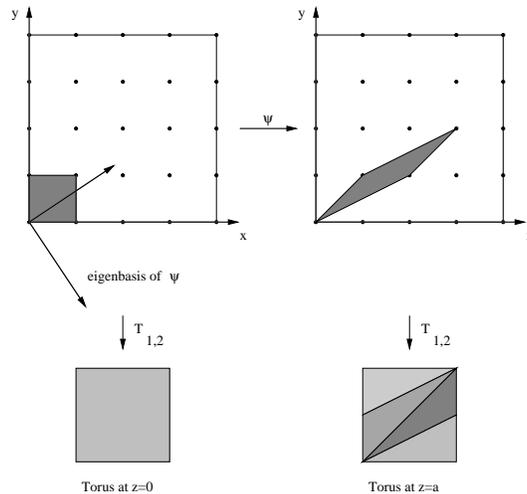}
\end{center}
\label{solveident}
\caption[Identifications of $\RR^{3}$ for a compact three-manifold 
modelled on solvegeometry.]
{Identifications of $\RR^{3}$ for a compact three-manifold 
modelled on solvegeometry, using $M=\left [ \begin{array}{cc} 2&1\\ 
1&1\end{array}\right ]$ in (\ref{eqII22}). The element $(0,0,a)$ of 
Sol maps rectangles on the left-hand face to parallelograms on the 
right, preserving the lattice of pairs of integers. The faces are 
each periodically identified, becoming tori.}
\end{figure}

Finally, the Sol solutions can be extended through the horizon in the 
usual way, but there will a curvature singularity at $r=0$ when $M\neq 0$; 
$R^{abcd}R_{abcd}\sim M^2/r^6$ there. The Hawking temperature,
computed from the regularity of the Euclideanized metric, is 
$(\Lambda^2M/ 24\pi^3)^{1/3}$, and so vanishes as $M\to 0$.

\subsection{Nilgeometry Black Holes}
For $M>0$, the $\Lambda<0$ Einstein metric
\begin{eqnarray}
ds^2 &=& -\left ( \frac{-2\Lambda}{11}r^2 - \frac{2M}{r^{5/3}}\right ) 
dt^2 + \frac{dr^2}{\frac{-2\Lambda}{11}r^2 - \frac{2M}{r^{5/3}}}\nonumber\\
&& + r^{4/3}\left ( dx^2 + dy^2 \right ) 
+ r^{8/3} \left ( dz - \sqrt{\frac{-4\Lambda}{9}}xdy\right )^2\label{eqII23}
\end{eqnarray}
has horizon located at $r=(11M/(-\Lambda))^{3/11}$. This horizon (and the
exterior spacetime) can be foliated by 3-manifolds modelled on nilgeometry.

For nilgeometry, we take the manifold $X$ to be the Heisenberg group, 
which is denoted Nil (it's a nilpotent Lie group) and consists of all 
$3 \times 3$ upper triangular matrices of the form
$\left[	\begin{array}{ccc} 
1 & x & z \\ 0 & 1 & y \\ 0 & 0 & 1\end{array}\right]$,
where $x,y,z \in \RR$. We can identify $(x,y,z) \in \RR^{3}$ with the above 
matrix, giving us the multiplication $(x,y,z)\cdot(x',y',z') = (x+x', y+y', 
z+z'+xy')$. We can determine a (left) invariant metric on $\RR^{3}$ under 
the action of Nil on itself by picking a metric arbitrarily at a point and 
using invariance. If we pick $ds^2 = dx^2 + dy^2 + dz^2$ at the origin, the 
resulting invariant metric is 
\begin{equation}
ds^2 = dx^2 + dy^2 + (dz - xdy)^2\quad .\label{eqII24}
\end{equation}
As pointed out in \cite{Fa1}, this metric is of Bianchi type~II. 
The rescalings $x \mapsto r^{2/3}x$, $y \mapsto r^{2/3}y$, $z \mapsto 
r^{4/3} z$ give the metric induced by (\ref{eqII23}) 
on each of the $t=const.$, $r=const.$\ surfaces.

As before, the metric (\ref{eqII23}) can be taken to describe a black 
$3$-brane with horizon $X$, but once again we can use isometries to 
compactify $X$ and obtain a black hole. A standard example of a compact 
$3$-manifold with geometric structure modelled on Nil can be constructed 
by taking the quotient of Nil by the subgroup $\Gamma$ of Nil consisting 
of matrices with only integer entries. If $a,b,c \in \ZZ$, then 
\begin{equation}
\left[	\begin{array}{ccc}
1 & a & c \\
0 & 1 & b \\
0 & 0 & 1
	\end{array}
\right] 
\left[	\begin{array}{ccc}
1 & x & z \\
0 & 1 & y \\
0 & 0 & 1
	\end{array}
\right]
=
\left[	\begin{array}{ccc}
1 & x + a & z + c + ay \\
0 & 1 & y + b \\
0 & 0 & 1
	\end{array}
\right]\quad .\label{eqII26}
\end{equation}
Thus, two points $(x,y,z)$ and $(x',y',z')$ are identified in 
$\textrm{Nil}/\Gamma$ iff $x'-x=a$, $y'-y=b$, and $z'-z=ay+c$
for any triple of integers $(a,b,c)$. Choosing $a=0$, we see 
that we should identify points within planes of constant $x$, 
turning these planes into tori. The resulting ``toroidal cylinder'' 
can be compactified using the $a\neq 0$ identifications. To 
identify the torus at $x$ with the torus at $x'=x+a$, for $a$ 
a non-zero integer, we take lines $z=k=const$, on the $yz$-plane 
(covering the torus) at $x$ and identify them with $z'=k+ay$ 
at $x'$. The identifications are shown in Figure~2. As with Sol,
it is an easy exercise to show that $\Gamma$ acts freely and 
properly discontinuously. The resulting manifold $\textrm{Nil}/\Gamma$
is a circle bundle over the torus.
\begin{figure}
\begin{center}
\includegraphics{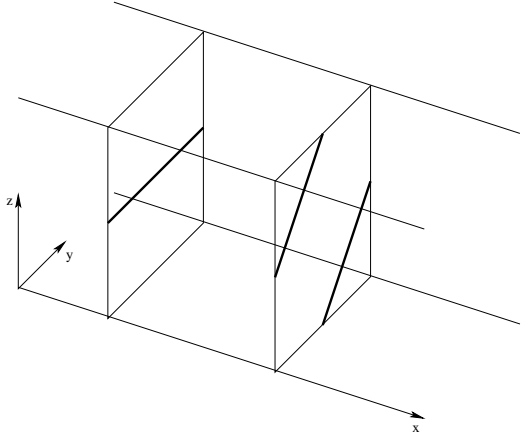}
\end{center}
\label{nilident}
\caption{Identifications of $\RR^{3}$ for a compact three-manifold 
modelled on nilgeometry  (based on Fig.\ 3.26 of \cite{Th1}). Opposite
sides of the square cylinder are identified. The displayed left- and 
right-hand faces are identified by matching up the lines.}
\end{figure}

The Nil solutions can be extended through the horizon but, for $M\neq 0$,
$r=0$ is a curvature singularity; $R^{abcd}R_{abcd}\sim M^2/r^{22/3}$ there. 
The Hawking temperature is $(11\Lambda^{8/3}M)^{3/11}/ 6\pi$, 
and so vanishes as $M\to 0$.\footnote
{In fact, of all solutions in this paper, the Hawking temperature goes 
to zero on approach to the ``ground state'' for all but the Kottler, 
Schwarzschild, and black string ($\SSS^2\times \SSS^1$) cases. In those cases, 
it diverges.}

\setcounter{equation}{0}
\section{A Bianchi VI$_{-1}$ Braneworld}
In this section, we will construct a simple ``braneworld'' cosmology 
by embedding a timelike hypersurface in each of two solvegeometry 
spacetimes. The braneworld construction involves cutting each 
spacetime along the hypersurface and glueing together one piece 
from each spacetime along the boundaries resulting from the cutting. 
One obtains by this process a spacetime with a singular hypersurface 
of induced metric $h_{ab}$ and extrinsic curvature $K_{ab}$ and a 
solvegeometry metric ``in the bulk'' ({\it i.e.}, everywhere else). 
The induced metric in the present case can be read off from 
(\ref{eqA11}) and (\ref{eqII21}):
\begin{equation}
h_{ab}dx^a dx^b = -d\tau^2 + f^2(\tau)\left ( e^{2z}dx^2 + e^{-2z}dy^2
\right ) +dz^2\quad ,\label{eqIII1}
\end{equation}
where the function $f(\tau)$ will be determined from junction conditions.
In some approaches, the bulk metric may differ on either side of this
hypersurface, because of different values of $M$. We will allow this,
and will distinguish the metric potential $V$ on the two sides as $V^+$ 
and $V^-$, resp. In any case, the hypersurface will be a singularity 
of the curvature, so by the junction conditions \cite{Is1} it has a 
surface energy density tensor $S^a{}_{b}$. For simplicity, we take 
$S^{\mu}{}_{\nu}$ to be diagonal in the coordinate system of the 
Appendix with eigenvalues $-\rho$, $p_x$, $p_y$, and $p_z$. The 
conservation law (\ref{eqA16}) gives
\begin{eqnarray}
&&\frac{\partial p}{\partial x}=\frac{\partial p}{\partial y}
=\frac{\partial}{\partial z}\left (p-\frac{1}{2}\rho\right ) 
=0\quad ,\label{eqIII2}\\
&&\frac{\partial}{\partial\tau}\left ( \rho f^2(\tau)\right )
=-p\frac{\partial}{\partial\tau}\left ( f^2(\tau)\right )\quad .
\label{eqIII3}
\end{eqnarray}

We now compute from (\ref{eqA13}, \ref{eqA14}, \ref{eqA15}) 
the junction conditions in the case that the bulk spacetime
is comprised of solvegeometry pieces. We get
\begin{eqnarray}
&{\rm {\mathit {uu}}-component:}&
\frac{d}{d\tau}\left [ 
\sqrt{V^+(f(\tau))+{f'}^2(\tau)}-\sqrt{V^-(f(\tau))+{f'}^2(\tau)}
\right ] \nonumber\\
&&= \frac{8\pi}{3}\left ( 2\rho+p_x+p_y+p_z\right )f'(\tau)
\label{eqIII4}\\
&{\rm {\mathit{xx}}-component:}&
\sqrt{V^+(f(\tau))+{f'}^2(\tau)}
+\sqrt{V^-(f(\tau))+{f'}^2(\tau)}\nonumber\\
&&= \frac{8\pi}{3}\left (2p_x-p_y-p_z
+\rho\right )
\label{eqIII5}\\
&{\rm {\mathit{yy}}-component:}&
\sqrt{V^+(f(\tau))+{f'}^2(\tau)}
+\sqrt{V^-(f(\tau))+{f'}^2(\tau)}\nonumber\\
&&= \frac{8\pi}{3}\left (2p_y-p_z-p_x
+\rho\right )
\label{eqIII6}\\
&{\rm {\mathit{zz}}-component:}&
0 = 2p_z - p_x - p_y +\rho
\label{eqIII7}
\end{eqnarray}
From the spatial component equations, we obtain
\begin{eqnarray}
p:=p_x=p_y=p_z+\frac{1}{2}\rho&&\quad ,\label{eqIII8}\\
\sqrt{V^+(f(\tau))+{f'}^2(\tau)}+\sqrt{V^-(f(\tau))+{f'}^2(\tau)} 
&=& 4\pi\rho f(\tau)\quad .\label{eqIII9}
\end{eqnarray}
It can be useful to re-express this last equation in the form
\begin{equation}
\sqrt{V^{\pm}(f(\tau))+{f'}^2(\tau)}=\pm\frac{\left ( V^+(f(\tau))
- V^-(f(\tau)) \right )}{8\pi\rho f(\tau)}+2\pi\rho f(\tau)\quad .
\label{eqIII10}
\end{equation}
In particular, (\ref{eqIII10}) can be used to cast (\ref{eqIII4}) in 
the form
\begin{equation}
\frac{d}{d\tau}\left [ \frac{V^+(f(\tau))- V^-(f(\tau))}
{4\pi\rho f(\tau)}\right ] 
=8\pi\left ( p+ \frac{1}{2}\rho \right ) f'(\tau)\quad ,\label{eqIII11}
\end{equation}
where we have also used (\ref{eqIII8}) to simplify the right-hand side.

We now analyze (\ref{eqIII9}) and (\ref{eqIII11}), beginning with the 
latter. By inserting the explicit form of $V(a)$ for the solvegeometry
black hole into the left-hand side, (\ref{eqIII11}) becomes
\begin{eqnarray}
8\pi\left ( p+\frac{1}{2}\rho\right ) f'(\tau) 
&=&\frac{d}{d\tau}\left ( \frac{-\Delta M}{2\pi\rho f^2(\tau)} 
\right )\nonumber\\
&=&\frac{-\Delta M}{\pi}\frac{p}{\rho^2 f^3(\tau)}f'(\tau)
\quad ,\label{eqIII12}
\end{eqnarray}
where in the last step we used the energy conservation law (\ref{eqIII3}),
and we have defined $\Delta M:=M^+-M^-$. This equation leads to three 
possibilities: 
\begin{itemize}
\item[({\it i})] We are in the steady state $f'(\tau)=0$.
Steady state solutions arise even in standard Friedmann cosmology, 
and are associated with non-uniqueness of solutions.
\item[({\it ii})] We have the equation of state $\rho=-2p$, whence 
$\Delta M=0$ and we have reflection symmetry in the ``bulk'' about 
the braneworld. We can now easily integrate the conservation law to 
get the result $\rho\propto 1/f(\tau)$. This model obeys the Weak and 
Dominant Energy Conditions, but not the Strong Energy Condition.
\item[({\it iii})] We have a rather complicated equation of state, 
typical of braneworlds, which we express below in the implicit form
\end{itemize}
\begin{equation}
f(\tau) = \left ( \frac{-p\Delta M}{2\pi\rho^2 (p+\rho/2)}\right )^{1/3}
\quad .\label{eqIII13}
\end{equation}

\noindent
For case ({\it ii}) above, the converse also holds: $\Delta M=0
\Rightarrow \rho=-2p$ (unless $f'(\tau)=0$). This restriction on the 
equation of state has no analogue in the more standard scenario of
reflection symmetry about a braneworld in a Kottler black hole bulk.
\footnote
{We thank Shinji Mukohyama for bringing this to our attention, and
for suggesting that this could be a consequence of our assumption of
a static bulk. His suggestion is that, since there is (apparently) no 
suitable analogue of the Birkhoff theorem for solvegeometry black 
holes, the requirement of a static bulk is non-trivial, and so can 
lead to non-trivial constraints on the braneworld \cite{Mu1}.}

Finally, we use the explicit form of $V(a)$ to recast (\ref{eqIII9}) 
in the form ``kinetic + potential energy = constant:''
\begin{eqnarray}
&&\frac{1}{2}{f'}^2(\tau)+U(f(\tau))=0\quad ,\label{eqIII14}\\
&&U(A)=\frac{1}{2}\left [ 1-\left ( \frac{\rho}{\rho}_c
\right )^2 \right ] \left (\frac{A}{\ell}\right )^2 -\frac{\left (
M^+ + M^-\right )}{2A}-\frac{(\Delta M)^2 \ell^2}
{128A^4}\left ( \frac{\rho_c}{\rho}\right )^2\quad ,\label{eqIII15}
\end{eqnarray}
where $\ell^2:=9/(-2\Lambda)$ and $\rho_c:= 2/(\pi\ell)$. 
The energy density $\rho$ enters (\ref{eqIII15}) as a square, as it 
does in FRW braneworld cosmology, but unlike in standard cosmology. 

As it is not our intention to explore the resulting cosmology in 
depth in the present work, we limit ourselves to a few observations
and assume, from here on, reflection symmetry $\Delta M=0$. As 
already noted, $\rho\propto 1/f(\tau)$ in this case, so we write
\begin{equation}
\rho(\tau)=\sigma\rho_c\ell/f(\tau)\quad ,\label{eqIII16}
\end{equation}
where $\sigma$ is a constant and use this to rewrite (\ref{eqIII14},
\ref{eqIII15}) as
\begin{equation}
{f'}^2(\tau)=\sigma^2-\left ( \frac{f(\tau)}{\ell}\right )^2
+\left ( \frac{M^++M^-}{f(\tau)}\right )\quad .\label{eqIII17}
\end{equation}
Somewhat remarkably, this is precisely the usual Friedmann equation for 
a negative spatial curvature (or flat, if $\sigma=0$), matter dominated, 
FRW cosmology with constant mass density $3(M^++M^-)/8\pi$, zero pressure, 
and cosmological constant $-3/\ell^2<0$. 

If $M^++M^-=0$, two kinds of solutions are well-known. There
is the steady state solution $f(\tau)=\sigma\ell$ and a family of 
solutions
\begin{equation}
f(\tau)=\sigma\ell\sin\frac{\tau-\tau_0}{\ell}\quad ,\label{eqIII18}
\end{equation}
where we should consider only one half-cycle, corresponding to the
birth of the universe at a singularity, its expansion to a maximum,
and its subsequent recollapse. At the maximum, the square root of the
right-hand side of \ref{eqIII17} is not Lipshitz. The uniqueness theorem
for solutions therefore fails \cite{ChDeDi1},\footnote
{This ``hesitant universe'' scenario is a well-known property of the
Friedmann equation. EW wishes to thank Connell McCluskey for a discussion 
on this point.}
so that a cosmos expanding to a maximum size can then remain in 
the steady state for an arbitrary time before recollapsing. When 
$M^++M^-\neq 0$, these qualititive details persist, though in this 
case the time coordinate $\tau$ diverges on approach to the singularity.

\setcounter{equation}{0}
\section{Discussion}
We have been able to give new solutions of the $5$-dimensional Einstein
equations. These solutions can be regarded as black branes with 
horizons given by solvegeometry and nilgeometry, resp., but standard 
techniques permit the horizons to be ``spatially compactified,''
turning the new solutions into black holes. As a result, static 
$5$-dimensional black holes are now known with horizons modelled on 
all but one of the Thurston model geometries.

The case of a horizon modelled on the $\SLR$ geometry remains open.
$\SLR$ is the universal cover of the group of $2 \times 2$ matrices 
with determinant $1$, and this model geometry corresponds to Bianchi 
type VIII.  $\SLR$ has the structure of a line bundle over the hyperbolic 
plane $\HH^{2}$, but is not isometric to the $\HH^{2} \times \RR$ case.
In this case, the field equations are somewhat more complicated than
the others and simple ans\"atze such as those we applied above do 
not work. At this stage, it is too early to say whether there is a 
fundamental obstruction or whether the difficulty is that our ans\"atze 
have been too na\"{\i}ve in this case. The issue remains under active 
consideration.

The new solutions raise several questions: are they subject to the 
Gregory-Laflamme instability \cite{GrLa1,Gr1} of black strings, can 
horizons of completely arbitrary topology be constructed in $5$ 
dimensions, and do any of these solutions have an interpretation 
in terms of compactifications of fundamental physics in higher dimensions? 
They also raise the following issue of black hole uniqueness. Assume 
specific fall-off behaviour near infinity, say that of locally asymptotically 
anti-de Sitter spacetimes. Will this fall-off behaviour and reasonable 
causality and energy conditions in the bulk spacetime preclude the existence
of a black hole with horizon modelled on Nil or Sol? This is a next 
logical step beyond the product spacetimes studied in this article.
Now the fundamental group of the domain of outer communications of this 
spacetime would be non-trivial, and then from topological censorship 
there comes the constraint that the fundamental group of scri must 
map onto it, but for hyperbolic scri this seems easy to arrange, and 
so spacetimes of this nature remain a possibility. To explore this 
issue, it may prove useful to consider whether there is a higher-dimensional 
analogue of Hawking's early approach \cite{Ha1,Ha2} to the question of black 
hole topology in $4$ dimensions.

\section*{Acknowledgements}
We would like to thank Shinji Mukohyama for correspondence 
about braneworlds. Our solutions were checked
using the GRTensorII \cite{grt} computer program.
This work was supported by a research grant from the 
Natural Sciences and Engineering Research Council of Canada.

\setcounter{equation}{0}
\appendix
\section{Braneworld Formalism}
In this appendix, we review braneworlds following the approach of
\cite{Kr1,Id2}, wherein the cosmos is a singular hypersurface
embedded in a five-dimensional spacetime. For any static metric of 
the form 
\begin{equation}
ds^2=-V(a)dt^2+\frac{da^2}{V(a)}+d\sigma^2(a,x^k)\label{eqA1}
\quad ,\end{equation}
where the coordinates $x^k$ parametrize the $t=const.$, $a=const.$ 
surfaces, consider an embedded timelike hypersurface parametrized 
by $(\tau,x^k)$ of the form
\begin{equation}
a=f(\tau)\quad , \quad t=g(\tau)\label{eqA2}\quad .
\end{equation}

At each point of the hypersurface, the vector
\begin{equation}
N^a:=\frac{f'(\tau)}{V(a)}\frac{\partial}{\partial t}+V(a)g'(\tau)
\frac{\partial}{\partial a}\label{eqA3}
\end{equation}
is normal to the hypersurface. We seek a vector field $n^a$ that is 
({\it i}) normalized and spacelike: $g(n,n)=1$, and
({\it ii}) tangent to an affinely parametrized geodesic 
congruence: $\nabla_nn=0$, that
({\it iii}) meets the hypersurface orthogonally: $n^a|_0
\propto N^a$, where $v|_0$ denotes the restriction of $v$ to
the hypersurface.
From ({\it i}) and ({\it ii}), and choosing the sign so that $n^a$ 
points to increasing $a$-values, we conclude that
\begin{equation}
n^a=\frac{E}{V(a)}\frac{\partial}{\partial t} +\sqrt{V(a)+E^2}
\frac{\partial}{\partial a}\quad ,\label{eqA4}
\end{equation}
where $E=E(\tau,x^k)$ is a constant of the motion along each integral 
curve of $n^a$, but depends on the parameters $(\tau,x^k)$ of the 
point at which the integral curve meets the hypersurface.
Condition ({\it iii}) determines $E$ such that the curve 
meets the hypersurface orthogonally, giving
\begin{equation}
E=f'(\tau)\sqrt{\frac{V(f(\tau))}{V^2(f(\tau)){g'}^2(\tau)
-{f'}^2(\tau)}}\quad .\label{eqA5}
\end{equation}
If we choose the parameter $\tau$ such that 
\begin{equation}
{g'}^2(\tau)=\frac{{f'}^2(\tau)+V(f(\tau))}{V^2(f(\tau))}\quad ,
\label{eqA6}
\end{equation}
then we obtain that
\begin{equation}
E= f'(\tau)\quad .\label{eqA7}
\end{equation}

We can read off $n^1:=da/d\lambda$ from (\ref{eqA4}) and integrate it 
using the fact that $E$ is constant along each integral curves of $n^a$, 
and so can express the affine parameter along these geodesics as
\begin{equation}
\lambda-\lambda_0 = \int\limits_{f(\tau)}^a
\frac{da'}{\sqrt{V(a')+E^2}}\quad .\label{eqA8}
\end{equation}

We now have a Gaussian normal coordinate system in a neighbourhood 
of the hypersurface, wherein each point $p$ is specified by coordinates 
$(\tau,\lambda,x^k)$ where $(\tau,x^k)$ are the parameter values at 
which the integral curve of $n^a$ that passes through $p$ meets the 
hypersurface, and $\lambda$ is the value of the affine parameter along 
this geodesic at $p$. That is, we can promote $\tau$ to a function 
on a neighbourhood of the hypersurface such that $\pounds_n \tau =
n^a\nabla_a \tau=0$. 
Rather than explicitly solving for $\tau$ in terms of $t$ and $a$, 
however, it suffices for our purposes to write the metric in the form
\begin{equation}
ds^2=-u\otimes u+d\lambda^2+d\sigma^2(a(\tau,\lambda),x^k)
\quad ,\label{eqA9}
\end{equation}
where
\begin{equation}
u_a=\sqrt{V(a)+E^2}dt-\frac{E}{V(a)}da\quad ,\label{eqA10}
\end{equation}
so $u^a:=g^{ab}u_b$ is unit past-timelike and orthogonal to $n^a$. 
Note that $[u,n]\neq 0$. By computing the restrictions of $dt$ and 
$da$ on the hypersurface, one can easily see that $u_a=d\tau$ there, 
so the first fundamental form $h_{ab}$ on the brane has line element
\begin{equation}
d{\tilde s}^2=h_{\mu\nu}dx^{\mu}dx^{\nu}
=-d\tau^2+d\sigma^2(f(\tau),x^k)\quad ,
\label{eqA11}
\end{equation}

This induced metric is not assumed to be governed by the 4-dimensional 
Einstein equation. Instead, the undetermined metric coefficients are
fixed by applying junction conditions to the second fundamental form.
If we denote the hypersurface as $\Sigma$, then for any $v^a,w^a\in 
T\Sigma$ and $n^a\in (T\Sigma)^{\perp}$ we have the second fundamental 
form
\begin{equation}
K(v,w)=v^aw^b\nabla_an_b
=\frac{1}{2}\left \{ \pounds_n\left ( g(v,w)\right ) + g(v,[w,n])
+g(w,[v,n])-g(n,[v,w])\right \}\label{eqA12}\quad .
\end{equation}

Now the procedure is to cut each of two copies of the 5-dimensional 
spacetime (designated here the ``$+$'' and ``$-$'' copies, resp.) 
along a timelike hypersurface with fundamental form $(h^{\pm}_{ab},
K^{\pm}_{ab})$, throwing away one side of each spacetime to create 
two spacetimes-with-timelike-boundary, and glueing these along the 
boundary. The result is a $5$ dimensional spacetime with a singular 
hypersurface where the glueing took place. This hypersurface is the 
{\it braneworld}. According to the junction conditions, it is necessary 
to prescribe a surface energy density along the boundary in order to 
balance the difference in the extrinsic curvatures of the two boundaries 
that were glued together \cite{Is1}.
\begin{equation}
K^+_{ab}-K^-_{ab}=-8\pi \left ( S_{ab}-\frac{1}{3} h_{ab}S \right )
\quad ,\label{eqA13}
\end{equation}
where $S_{ab}$ is the surface energy tensor, $S:=h^{ab}S_{ab}$, and
signs as superscripts distinguish the extrinsic curvatures on the two 
sides of the braneworld (we do not {\it a priori} assume a $\ZZ_2$
symmetry of reflection through the braneworld). In order that 
$\partial/\partial a$ match up across the braneworld, we must reverse 
it on one side, say the ``$-$'' side, since otherwise it would point 
outward from both sides of the braneworld. This leads to a $\pm$ sign 
in the spatial components of $K_{ab}$, which are
\begin{equation}
K^{\pm}_{ij}= \pm\frac{1}{2}n^a\partial_a h_{ij} 
= \frac{1}{2}\sqrt{V^{\pm}(f(\tau))
+{f'}^2(\tau)}\frac{\partial}{\partial a}h_{ij}|_{a=f(\tau)}\quad ,
\label{eqA14}
\end{equation}
The time-time component, here denoted $K_{uu}$, obeys
\begin{equation}
K^{\pm}_{uu}=\frac{-1}{f'(\tau)}\frac{d}{d\tau}
\sqrt{V^{\pm}(f(\tau))+{f'}^2(\tau)}\label{eqA15}\quad .
\end{equation}
The Gauss-Codazzi relations can be applied to the extrinsic curvature
to yield identities given in \cite{Is1}, including the conservation law
\begin{equation}
S^a{}_{b;a}=0\label{eqA16}\quad .
\end{equation}

\end{document}